\documentstyle[12pt,psfig]{article}
\def\reference{\parskip 0pt\par\noindent\hangindent 0.5 truecm}

\newcommand{\lapp}{\mbox{\raisebox{-0.3em}{$\stackrel{\textstyle <}{\sim}$}}}
\newcommand{\gapp}{\mbox{\raisebox{-0.3em}{$\stackrel{\textstyle >}{\sim}$}}}
\begin{document}
%
%
\title{Symmetry Parameters of CSS Sources: Evidence of Fuelling?}
%


\author{D.J. Saikia$^{1}$,
 S. Jeyakumar$^{2}$,
 F. Mantovani$^{3}$,
 C.J. Salter$^{4}$,  \\
 R.E. Spencer$^{5}$,
 P. Thomasson$^{5}$ and
 P.J. Wiita$^{6}$
} 

\date{}
\maketitle

{\center
$^{1}$ NCRA, TIFR, Post Bag No. 3, Ganeshkhind, Pune 411 007, India \\
djs@ncra.tifr.res.in \\[3mm]
$^{2}$ Physikalisches Institut, Universit\"{a}t zu K\"{o}ln, 50937 K\"{o}ln, Germany \\
jeyaks@ph1.uni-koeln.de \\[3mm]
$^{3}$ Istituto di Radioastronomia, CNR, Via P. Gobetti 101, I-40129 Bologna, Italy \\
fmantova@ira.cnr.it \\[3mm]
$^{4}$ Arecibo Observatory, HC3 Box 53995, Arecibo, PR 00612, USA  \\
csalter@naic.edu \\[3mm]
$^{5}$ Jodrell Bank Observatory, Macclesfield, Cheshire SK11 9DL, UK \\
pt@jb.man.ac.uk, res@jb.man.ac.uk \\[3mm]
$^{6}$ Georgia State University, Atlanta GA 30303-3083, USA \\
wiita@chara.gsu.edu \\[3mm]
}

%
\begin{abstract}
The compact steep spectrum (CSS) and gigahertz peaked spectrum (GPS) sources are widely believed to be young radio
sources, with ages $\lapp$10$^6$ yr. If the activity in the nucleus is fuelled by the supply of gas, one might
find evidence of this gas by studying the structural and polarisation characteristics of CSS sources and their
evolution through this gas. In this paper we discuss some of the possible `smoking-gun' evidence of this gas which
may have triggered and fuelled the radio source.
\end{abstract}


\bigskip

%
%

\section{Introduction}
There is a  consensus of opinion that the compact steep spectrum (CSS) and gigahertz peaked spectrum (GPS) sources
are young objects seen at an early stage of their evolution. Recent measurements of component advance speeds for a
few very compact sources yield ages of about 10$^3$ yr (Owsianik \& Conway  1998; Taylor et al.\ 2000). Spectral
studies of CSS sources
 suggest ages $\lapp$10$^5$ yr  (Murgia et al.\ 1999).
Models have been constructed for the evolution of extragalactic radio sources from the most compact symmetric
objects, christened CSOs, to the medium symmetric objects (MSOs) and later on to the standard FR\,II radio sources
(e.g.\ Carvalho 1985; Fanti et al.\ 1995; Readhead et al.\ 1996; O'Dea 1998 and references therein).  Numerical
simulations of jet propagation can reproduce these evolutionary steps (e.g.\ De Young 1997). Evolutionary
scenarios for CSOs constrained by their self-similar evolution (Jeyakumar \& Saikia 2000) and dependence of radio
luminosity on linear size (Snellen et al. 2000) have also been explored (Perucho \& Mart\'{\i} 2002).

An interesting related question is the fuelling of these young radio sources, possibly due to the infall of gas to
the central regions due to interacting companions and mergers. Detailed calculations  (e.g.\ Hernquist \& Mihos
1995) showed that following a merger the infall of interstellar matter into the central few hundred pc takes place
over a time scale of $\approx$10$^8$ yr. In such a situation, it should be possible to find evidence of this
infalling material by following the evolution of the radio components in these young radio sources as they advance
outwards and interact with this material.

\begin{figure}
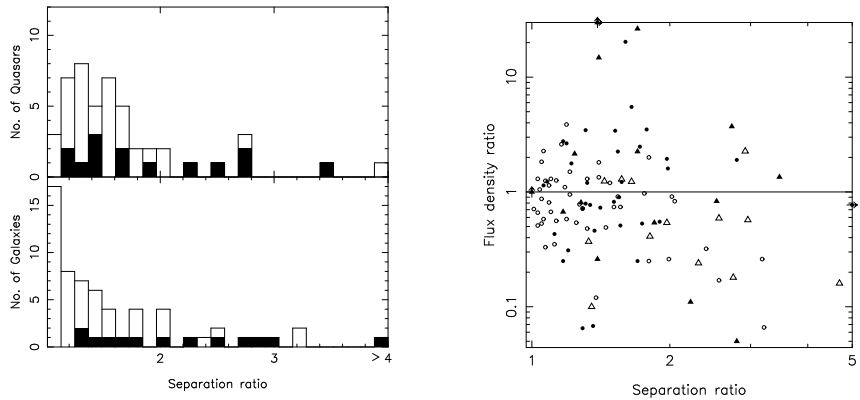

\hspace{1.5cm}
\begin{center}
\hbox{
\hspace{0.7in}
\psfig{file=fig1a.ps,width=2.0in}
\hspace{2.0in}
\psfig{file=fig1b.ps,width=2.0in}
}
\end{center}
\vspace{-1.05cm} \caption{Left: The distributions of the arm-length or separation ratio, r$_{\rm D}$, for the
sample of high-luminosity FR\,II radio galaxies and quasars. The CSS sources are shown in black. Right: The
r$_{\rm D}$--r$_{\rm L}$ diagram, where the filled and open symbols represent quasars and galaxies, respectively.
The CSS sources are shown as triangles. } \vspace{-.01cm}
\end{figure}

\section{Symmetry Parameters}
\subsection{Arm-Length and Flux Density Ratio}
One straightforward test is to examine the symmetry parameters, namely the arm-length or separation ratio, r$_{\rm
D}$, of the outer lobes from the nucleus, and the flux density ratio, r$_{\rm L}$, of the lobes. In order to
estimate reliably the separation ratio, r$_{\rm D}$, one also needs to identify the radio core or nuclear
component. Of the initially symmetric but oppositely-directed radio lobes, the one propagating in the direction of
more infalling material will encounter a denser ambient medium, and thus should be both brighter and closer to the
nucleus. However, this test is unlikely to be a sharp one because the infalling material or a merging galaxy is
likely to cover only a relatively small solid angle, depending on its size and separation from the nucleus. The
symmetry parameters would also depend on the orientation, bulk relativistic motion and evolution of the individual
components.

The distributions of r$_{\rm D}$, defined to be $>$\,1, for a sample of 109 FR\,II sources from the 3CRR and S4
samples of radio sources with a radio luminosity at 178\,MHz $\gapp10^{26}$\,W\,Hz$^{-1}$\,sr$^{-1}$ (H$_{\rm o}$
= 100\,km\,s$^{-1}$\,Mpc$^{-1}$ and q$_{\rm o}$ = 0) and a detected radio core are shown in Figure 1. A CSS source
is defined to be $<$20\,kpc in size. However, any limiting value is likely to be somewhat ad hoc, and one could
sometimes find evidence of interactions in larger sources as well. The CSS sources associated with both radio
galaxies and quasars have flatter distributions with higher median values, indicating their evolution in an
asymmetric environment (cf. Saikia et al. 2001). The median value of r$_{\rm D}$ for the CSS galaxies is $\sim2$
while the value for the larger galaxies is only 1.3. The corresponding values for the quasars are 1.7 and 1.5
respectively. The tendency for smaller sources to be more asymmetric was also reported by Arshakian \& Longair
(2000). The deficit of very symmetric quasars is due to their orientation close to the line of sight. In the
r$_{\rm D}$--r$_{\rm L}$ diagram for this sample (Figure 1, right), the nearer component is brighter for sources
with r$_{\rm L}$$<$1, and these have possibly encountered an asymmetric environment. Approximately 65\% of the
galaxies have r$_{\rm L}$$<$1 compared with $\sim$50\% for quasars, the difference being due to the smaller
orientation angles of quasars. Considering only those with r$_{\rm L}$$<$1, the median value of r$_{\rm L}$ is
$\sim$0.4 for CSS sources and close to $\sim$0.6 for the larger objects, suggesting again that the CSS sources are
evolving in a more asymmetric environment.

\subsection{Highly Asymmetric Individual Sources}
In this section, steep spectrum sources with large arm-length or separation ratios (r$_{\rm D}$$>$4) are
discussed. One of the most asymmetric CSS sources is the radio galaxy 3C\,459, with the ratio of separations of
the outer hot spots, r$_{\rm D}$, being $\sim$5 and the corresponding flux density ratio, r$_{\rm L}$, of the
oppositely-directed lobes being $\sim$0.45 and 0.3 at 5 and 1.7\,GHz respectively. High-resolution radio images of
this source have been produced recently by Thomasson, Saikia, \& Muxlow (2003). Their MERLIN+VLA image of the
source is shown in Figure 2. The eastern component, which is brighter, closer to the nucleus and more strongly
depolarised is interacting with denser gas. 3C\,459 has a young stellar population and high infrared luminosity,
and may have undergone a recent starburst, possibly triggered by the merging of a companion galaxy (Tadhunter et
al. 2002, and references therein). An HST image of the galaxy shows evidence of filamentary structures, which are
possibly of tidal origin, and a prominent peak of emission close to that of the nucleus of this N-galaxy, which
could be due to a galaxy in the late stages of merging (Thomasson et al. 2003).

From the compilation of symmetry parameters of high-luminosity 3CRR and S4 sources (Figure 1), there are only two
objects with a separation ratio $>$4, namely 3C\,254 and the CSS source B0428+205. 3C\,254 is associated with a
quasar at a redshift of 0.734 with r$_{\rm D}$=6.95 and r$_{\rm L}$=0.77 (Owen \& Puschell 1984; P.\ Thomasson et
al., in preparation). Optical line and continuum imaging of this source shows an extended emission line region
with the lobe on the nearer side interacting with a cloud of gas (Bremer 1997; Crawford \& Vanderriest 1997).
Although the flux density of the nearer lobe is brighter, the ratio is modest, which could be a consequence of
relativistic beaming of the hot spot further from the nucleus. The CSS object B0428+205, which has a largest
angular size of only 250\,mas, is associated with a galaxy at a redshift of 0.219 with r$_{\rm D}$=4.69 and
r$_{\rm L}$=0.16 (Dallacasa et al. 1995), consistent with a high dissipation of energy on the side where the jet
interacts with a dense cloud. Another example of a source with such a high degree of positional asymmetry is the
quasar 3C\,2 at a redshift of 1.037 for which r$_{\rm D}$=4.7 and r$_{\rm L}\sim$0.25 (Saikia, Salter, \& Muxlow
1987).  There is an indication of a radio jet extending from the core to the northern lobe. Depolarisation
gradients in both the lobes suggest interaction with the external medium, but the northern lobe seems to be more
significantly affected.

\begin{figure*}
\begin{center}
\psfig{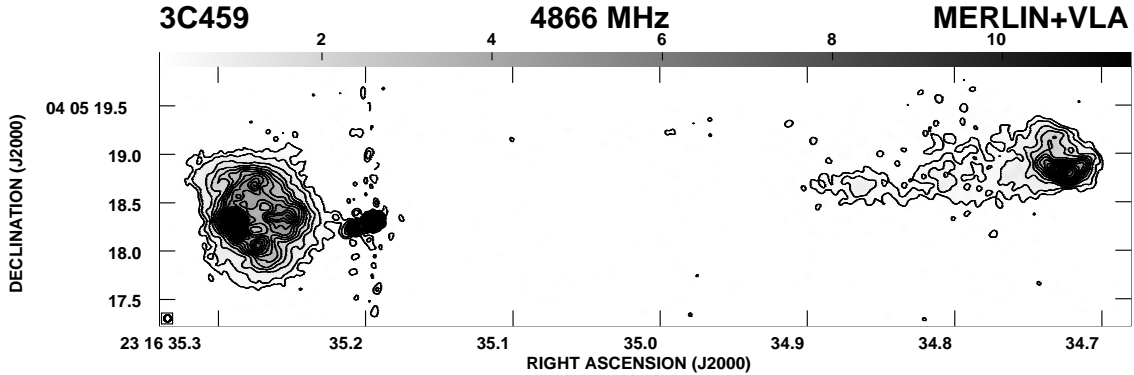}
\end{center}
\vspace{-2.4cm} \caption{The MERLIN+VLA image of 3C\,459 at 4866\,MHz with an angular resolution of 70\,mas. Peak
brightness: 391\,mJy/beam; contours: 0.29$\times$($-$1, 1, 2, 4, 6, 8, 10 $\ldots$ 36, 38, 40)\,mJy/beam. }
\end{figure*}

In the compilation of McCarthy, van Breugel, \& Kapahi (1991), there are only four galaxies with r$_{\rm D}>$ 4,
namely, 3C\,99, 3C\,208.1, 3C\,459, and 3C\,460. All these sources exhibit a large asymmetry in their flux density
ratio, with the brighter lobe being closer to the nucleus. In the case of 3C\,460, where they detect significant
extended emission line gas, the surface brightness of this gas is much higher on the side of the lobe closer to
the nucleus.  The radio galaxy 3C\,99 is similar to 3C\,459 in that it also has a steep spectrum radio core which
was resolved by the EVN and shows multiple components. Its value of r$_{\rm D}$=4.8, while r$_{\rm L}\sim$0.04 and
0.03 at 5 and 1.7\,GHz respectively, is consistent with the possibility that the jet on the side of the nearer
lobe is interacting with denser gas. Optical spectroscopic observations along the axis of the source do indeed
show that the gas on the side of the nearer component is blueshifted while that on the opposite side is redshifted
relative to the galaxy. The blueshifted gas is possibly being pushed towards us by interaction with the radio jet
(Mantovani et al. 1990).

\subsection{Flux Density Asymmetry for a Larger Sample}
The flux density ratio alone could be examined for a larger sample of sources since here one does not require the
detection of a radio core. Towards this end, a sample of sources of intermediate strength selected from the
408\,MHz Bologna B2.1 catalogue (Colla et al. 1970) has been observed with the VLA A--array (Saikia et al. 2002).
The sample consists of 52 objects which satisfy the following criteria. They lie within the region, 23$^{\rm h}$
30$^{\rm m}$ $<$ RA(B1950) $<$ 2$^{\rm h}$ 30$^{\rm m}$ and 29$^\circ$ 18$^\prime$ $<$ dec(B1950) $<$ 34$^\circ$
02$^\prime$, with their flux densities in the range 0.9$\leq$S$_{408}$$<$2.5\,Jy. Of these 52, 19 candidate CSS
sources have been observed with the VLA A--array at 4835\,MHz. A few of the radio images are shown in Figure 3. In
addition to this sample, the CSS sources from the B3+VLA sample observed by Fanti et al.\ (2001) and the 3CRR
sources have also been considered.

\begin{figure*}[h]
\begin{center}
\hspace{0.2cm}
\hbox{
  \hspace{0.5cm}
  \vbox{
  \psfig{file=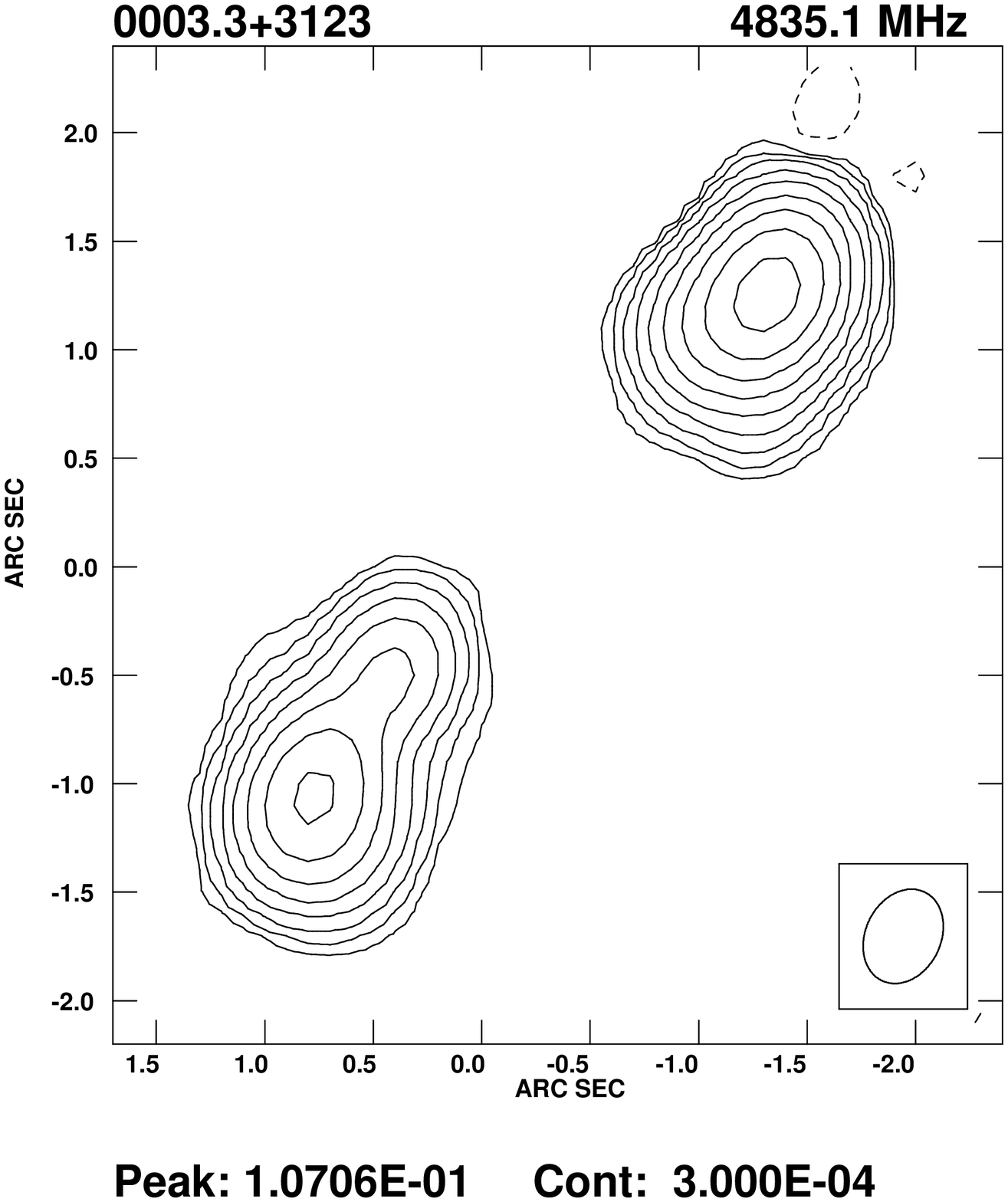,width=1.5in}
  \psfig{file=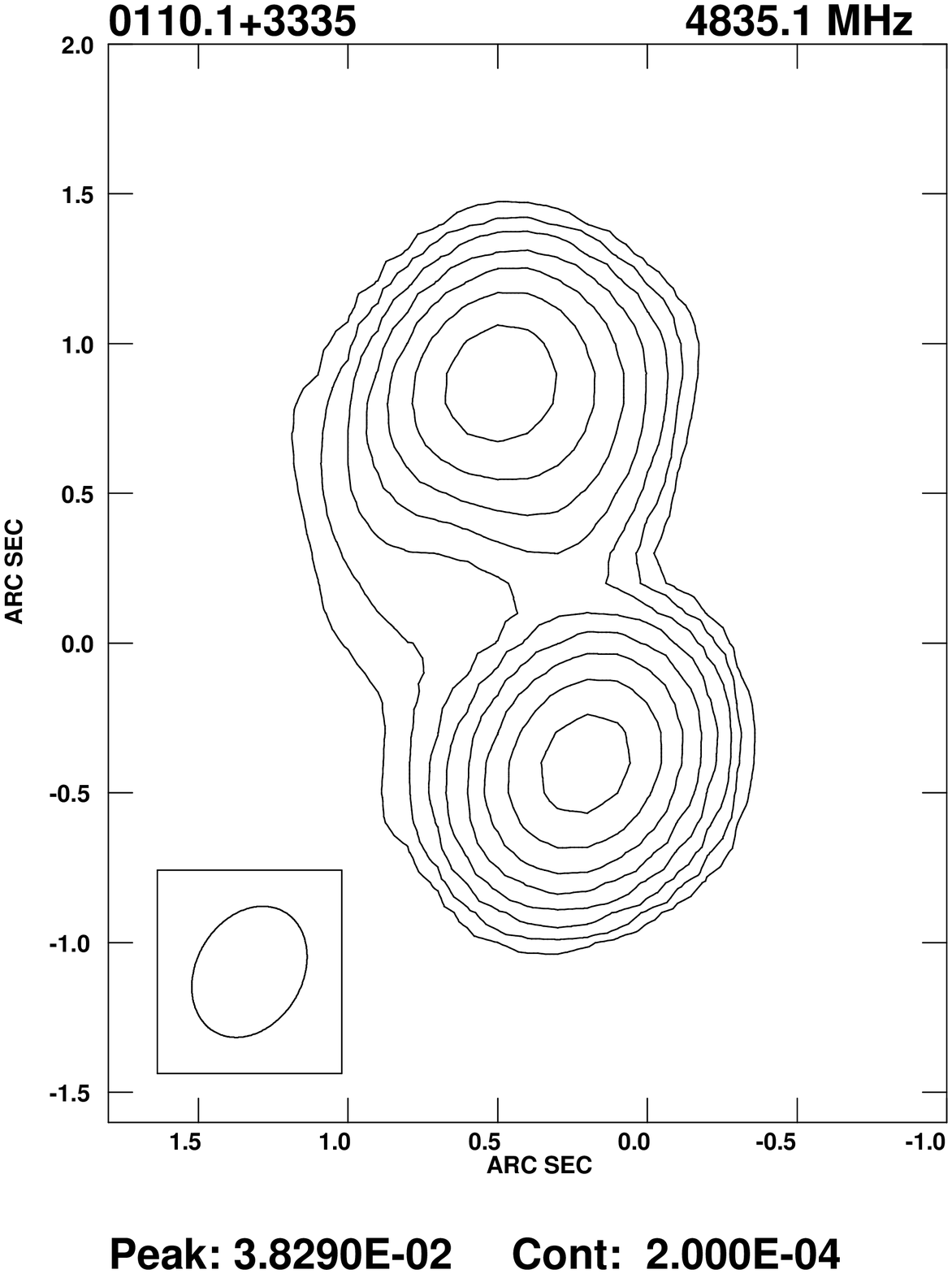,width=1.5in}
  }
  \vbox{
  \psfig{file=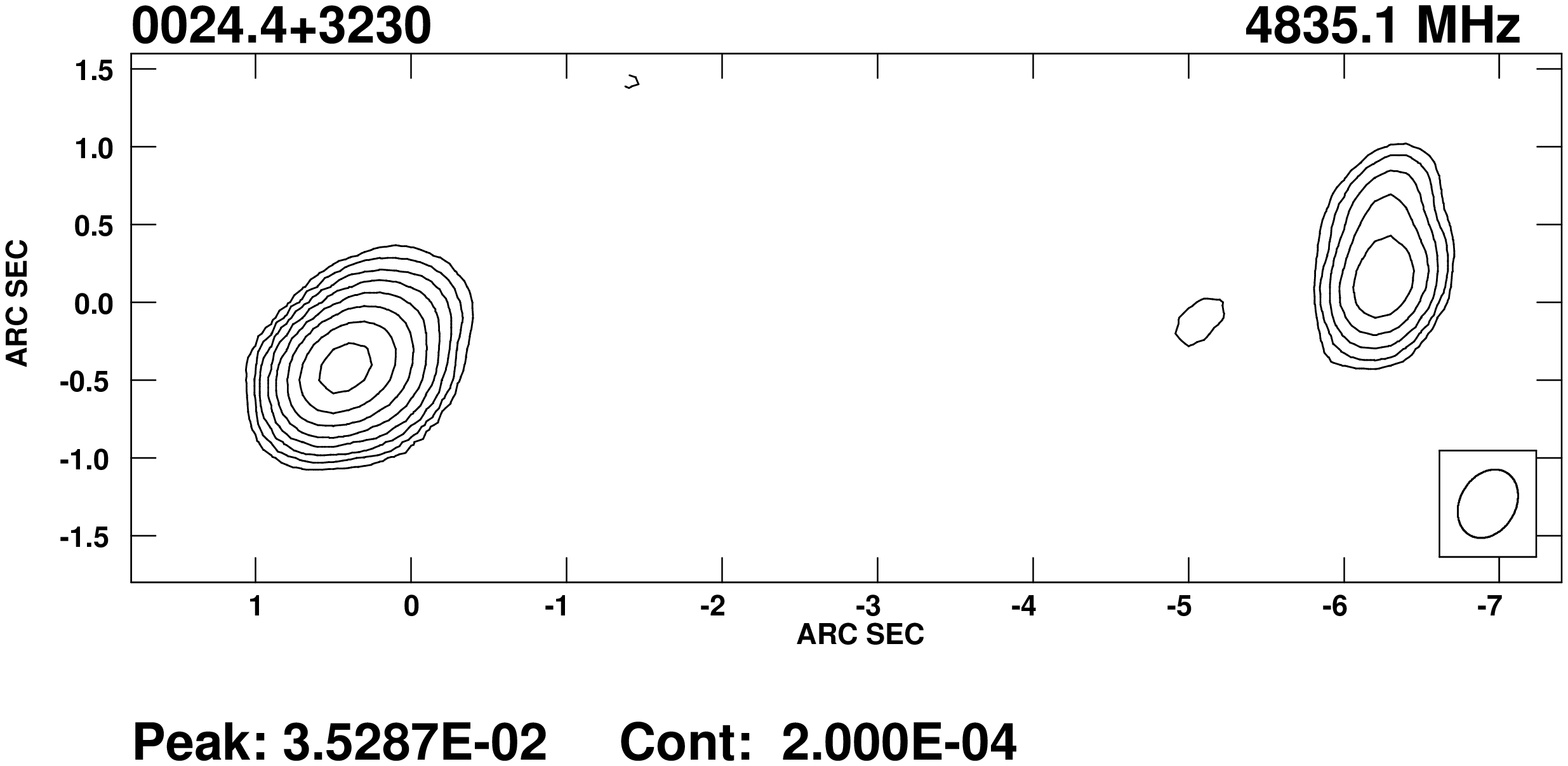,width=1.8in,angle=-90}
  \psfig{file=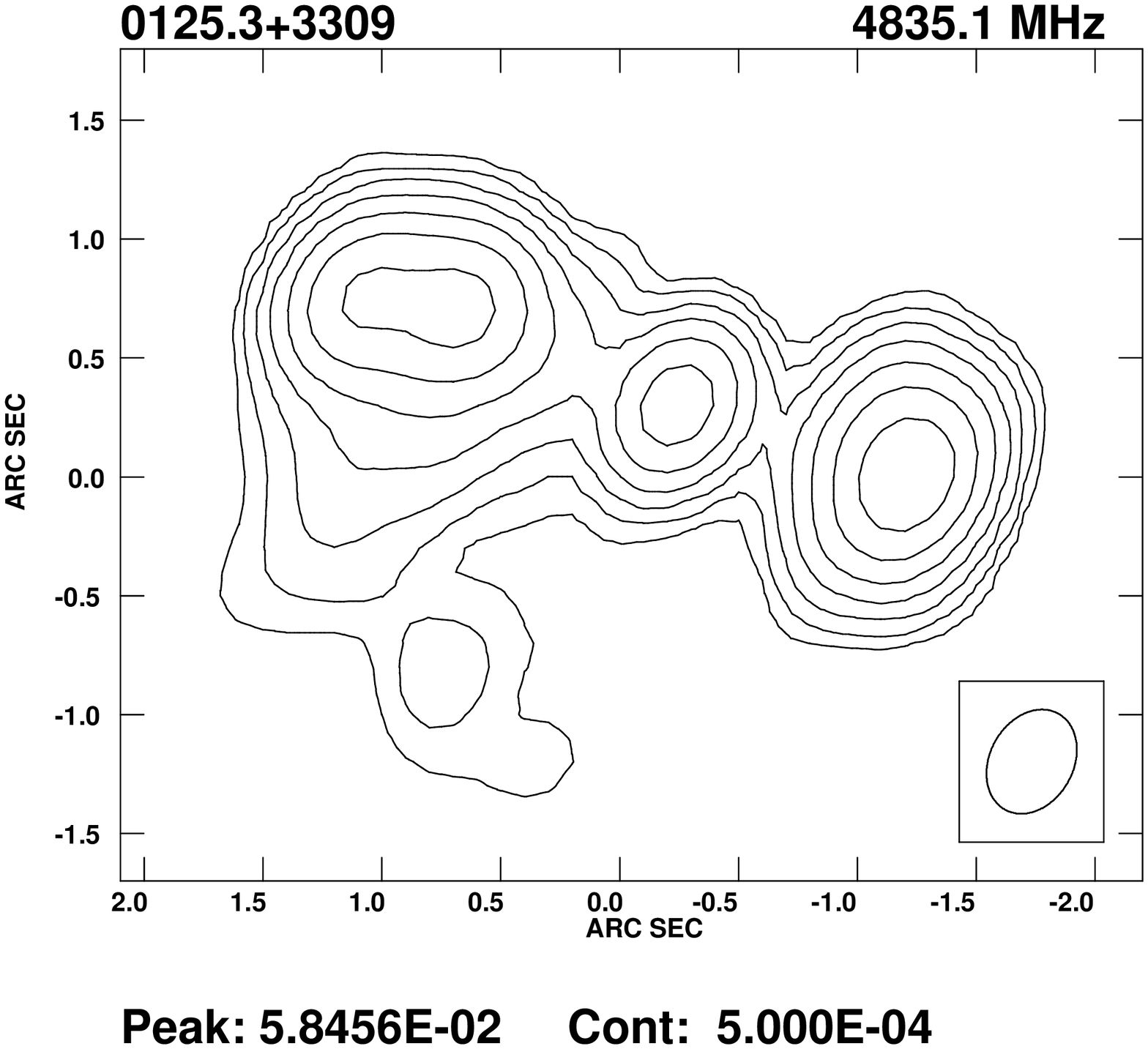,width=1.5in,angle=-90}
  \psfig{file=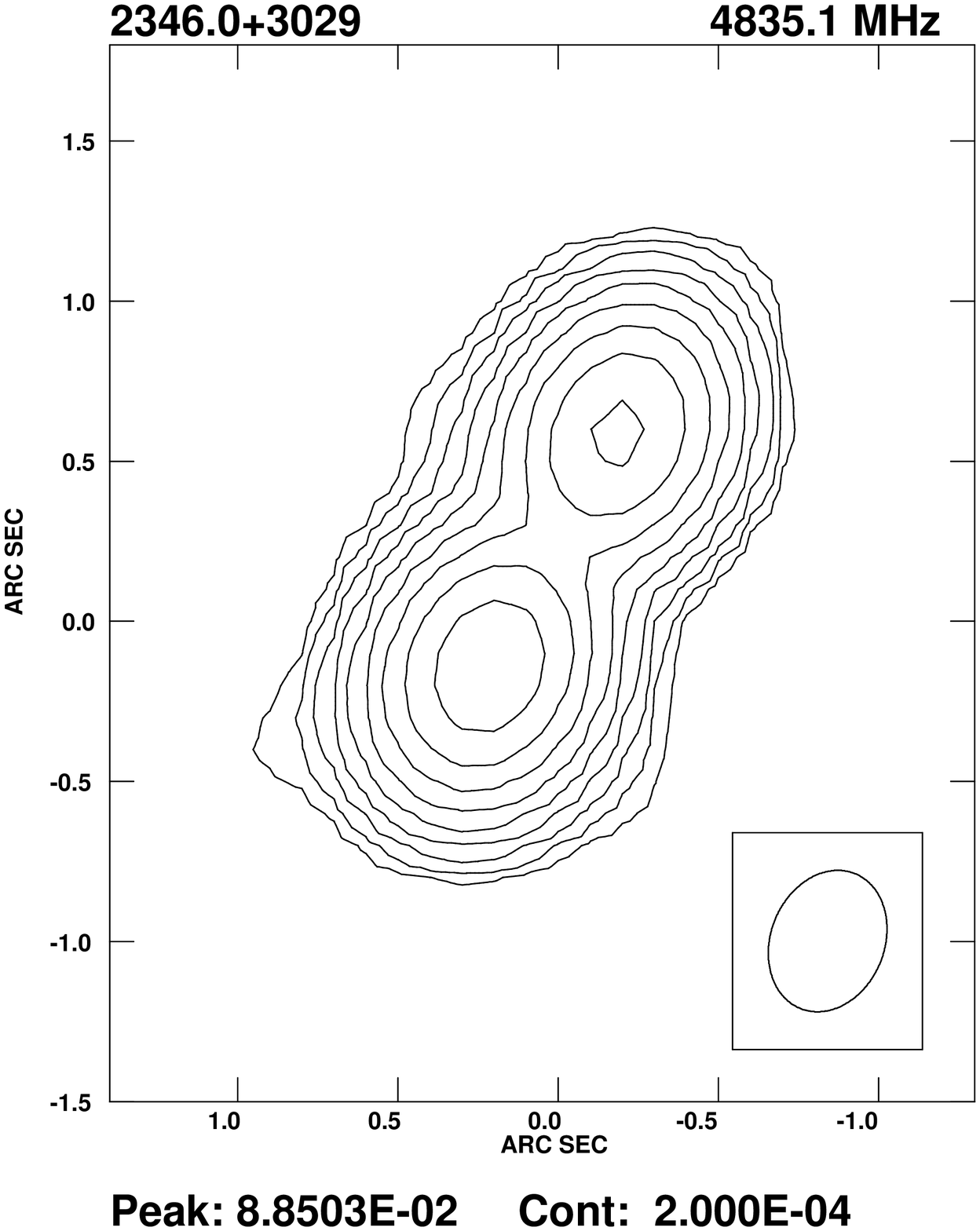,width=1.5in}
  }
  \vbox{
  \psfig{file=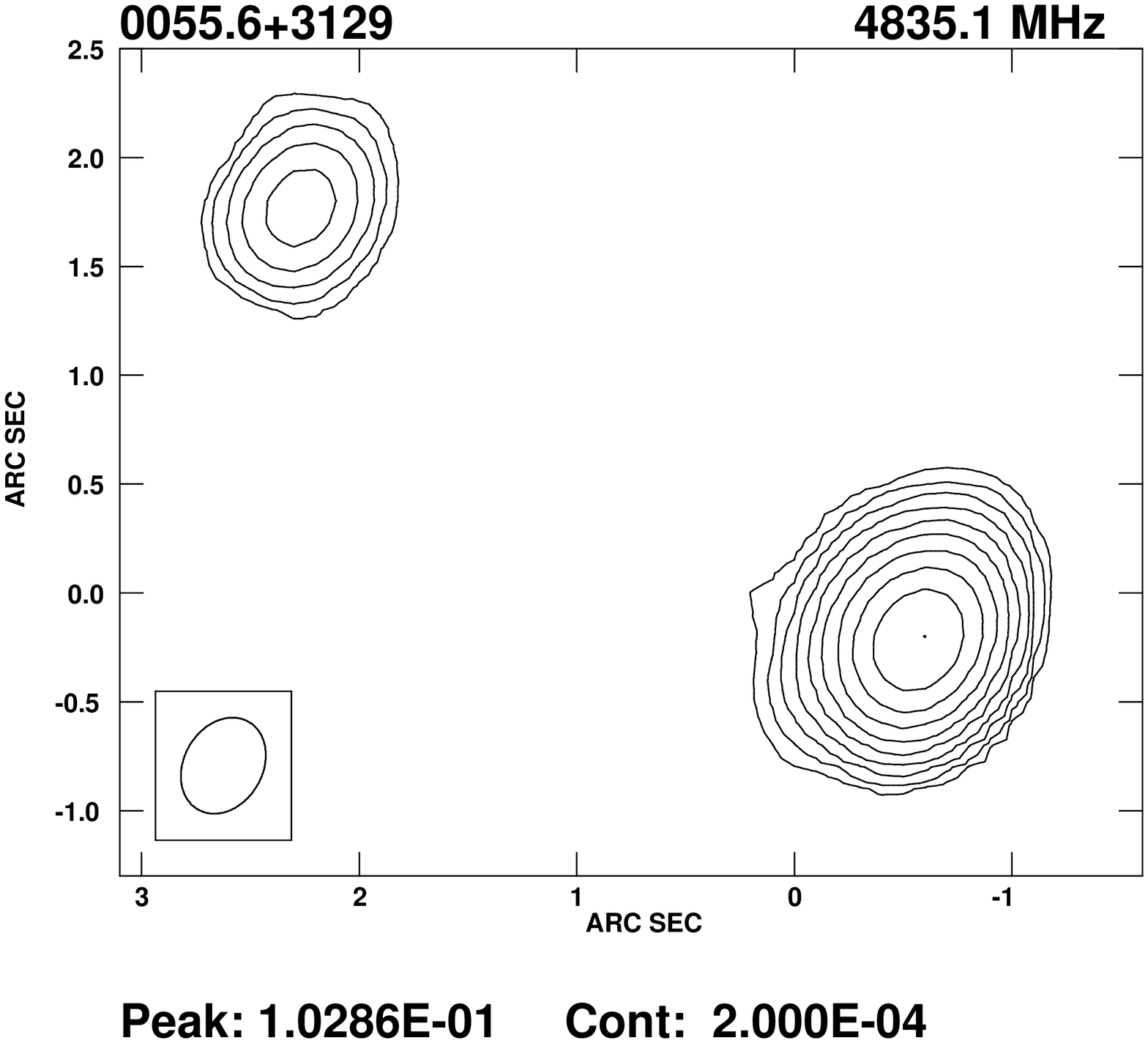,width=1.5in,angle=-90}
  \psfig{file=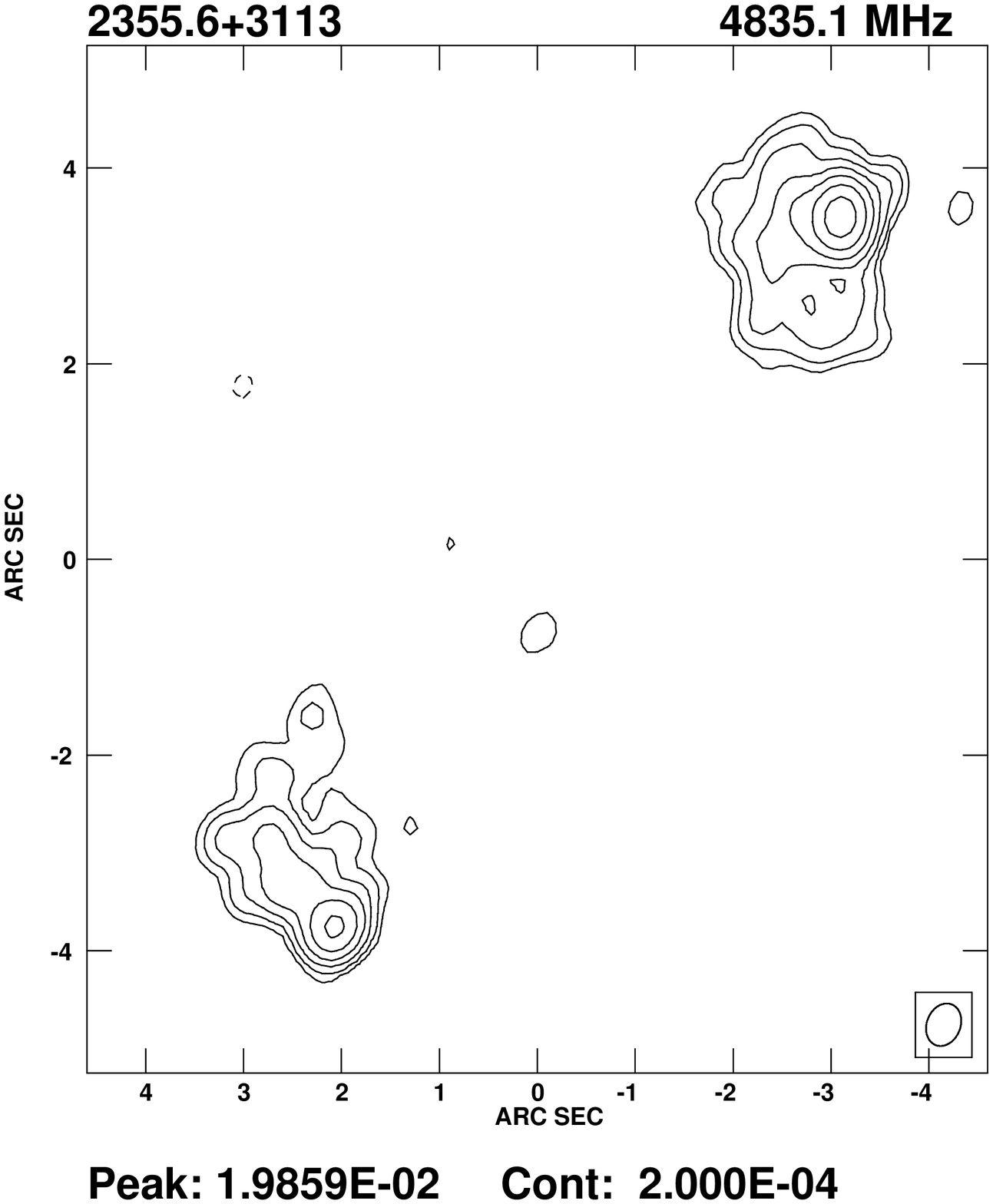,width=1.5in}
  }
} \caption[]{VLA A--array images of some of the candidate CSS objects from the B2 sample of sources. The contour
levels for all the images are $-$1, 1, 2, 4, 8, 16, \ldots times the first contour level. The peak brightness in
the image and the level of the first contour in units of Jy/beam are given below the images. The half-power
ellipse of the restoring beam is shown in one of the lower corners of each image. }
\end{center}
\end{figure*}

\noindent {\bf The B2 sample:} Considering the steep spectrum, double-lobed B2 sources, the `small-source' sample
is defined to consist of those with an LAS $<$10\,arcsec. The remaining objects constitute the `large-source'
sample. The median angular size of the `small-source' sample is $\sim$4\,arcsec compared with $\sim$40\,arcsec for
the larger sources. The flux density ratio of the oppositely directed lobes, r$_{\rm L}$, has been estimated from
either our observations or from those in the literature.  r$_{\rm L}$ ranges from $\sim$1.4 to 20 for the
small-source sample, and has a median value of $\sim$2, while for the large-source sample, r$_{\rm L}$ ranges from
$\sim$1.0 to 7, with a median value of $\sim$1.4 (Figure 4, left).

A significant fraction of the small sources show large asymmetries. Defining a very asymmetric source to be one
with r$_{\rm L}$ $>$ 5, we find that 5 of the 12 small-sized objects ($\sim$40\%) are highly asymmetric, while
only one of the 22 ($\sim$5\%) objects in the large-source B2 sample has r$_{\rm L}$ $>$ 5.

\begin{figure}
\hspace{2.5cm}
\hbox{
\vbox{
  \psfig{file=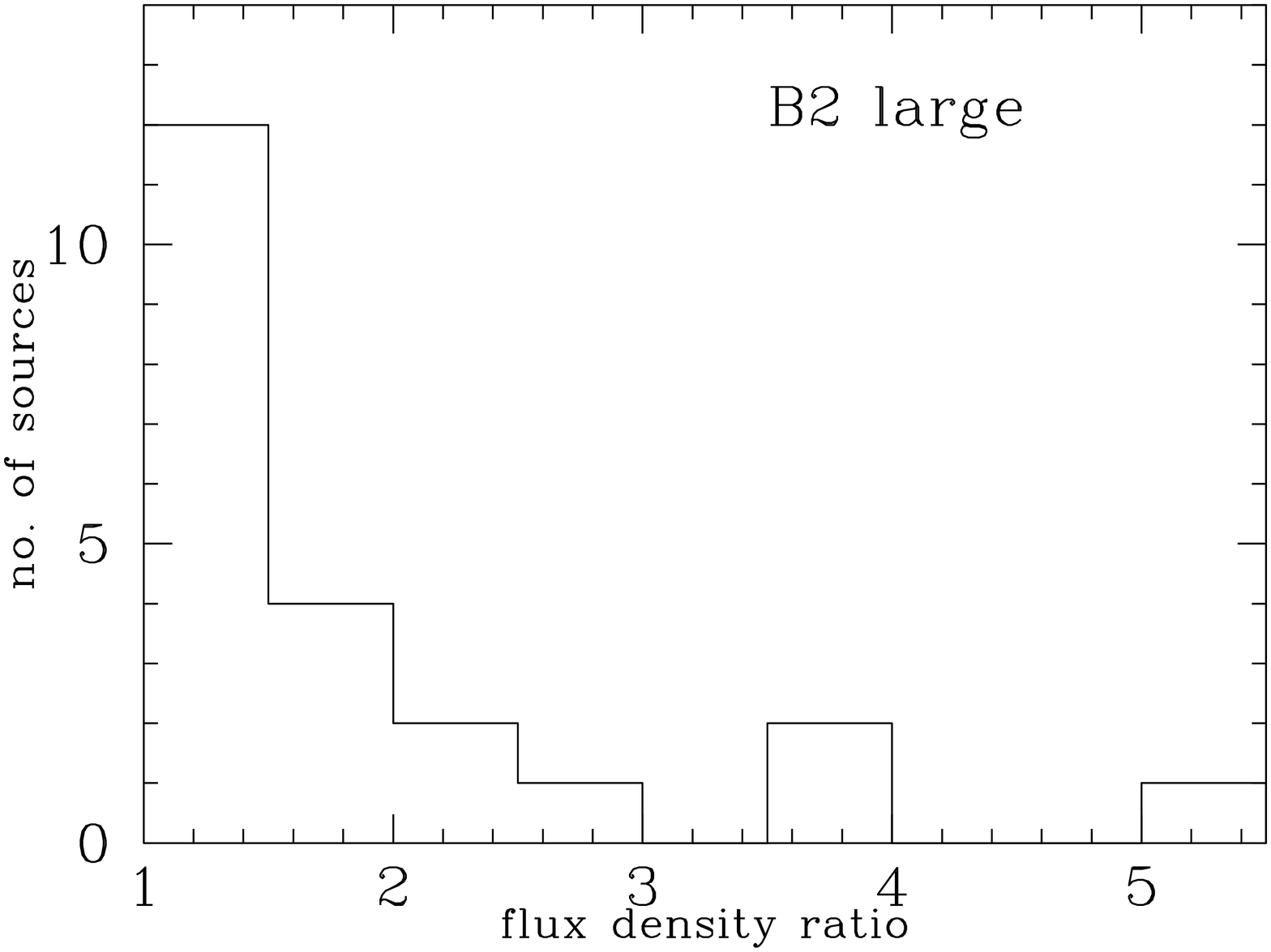,width=1.75in,angle=-90}
  \psfig{file=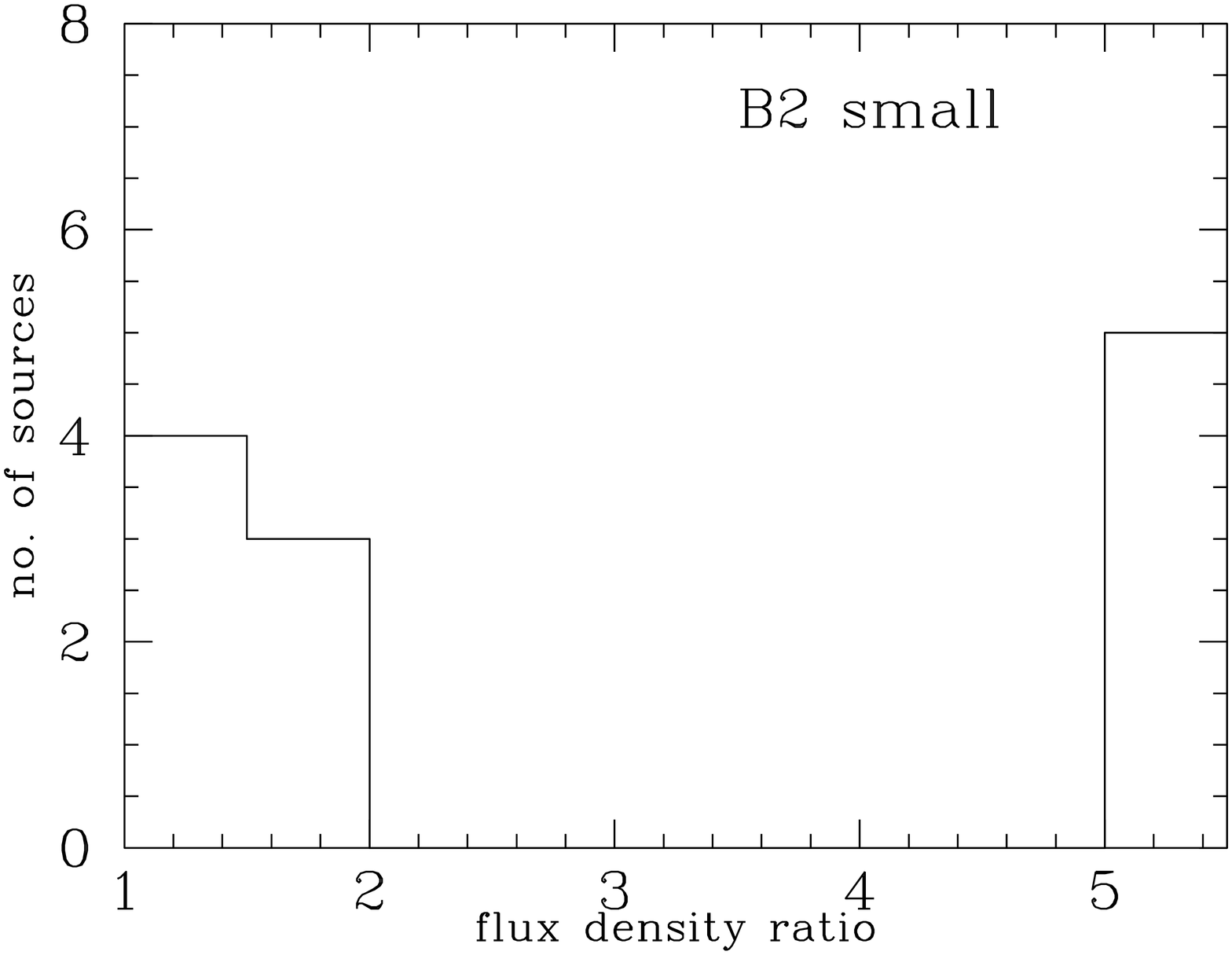,width=1.75in,angle=-90}
   }
\vbox{
  \psfig{file=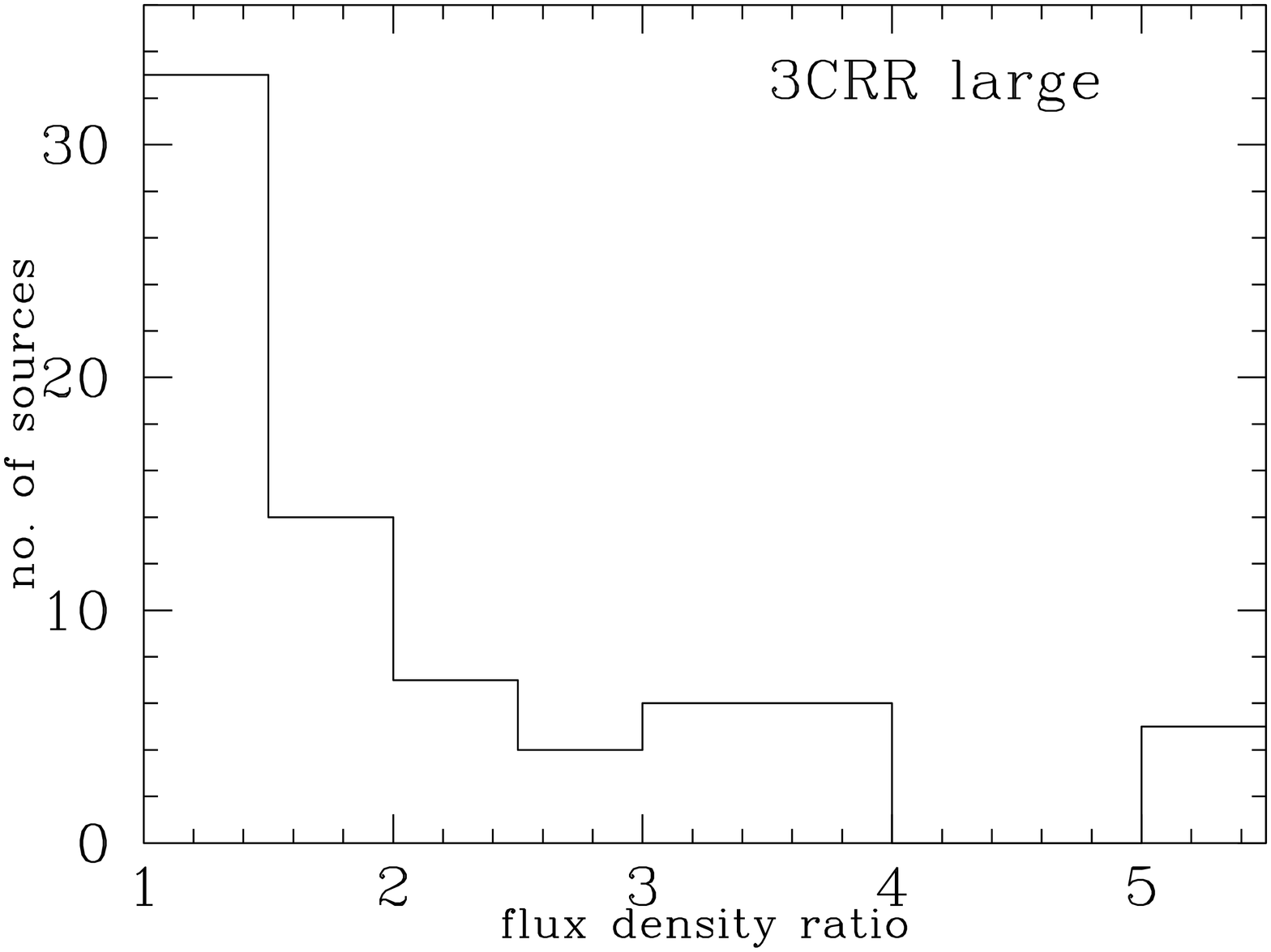,width=1.75in,angle=-90}
  \psfig{file=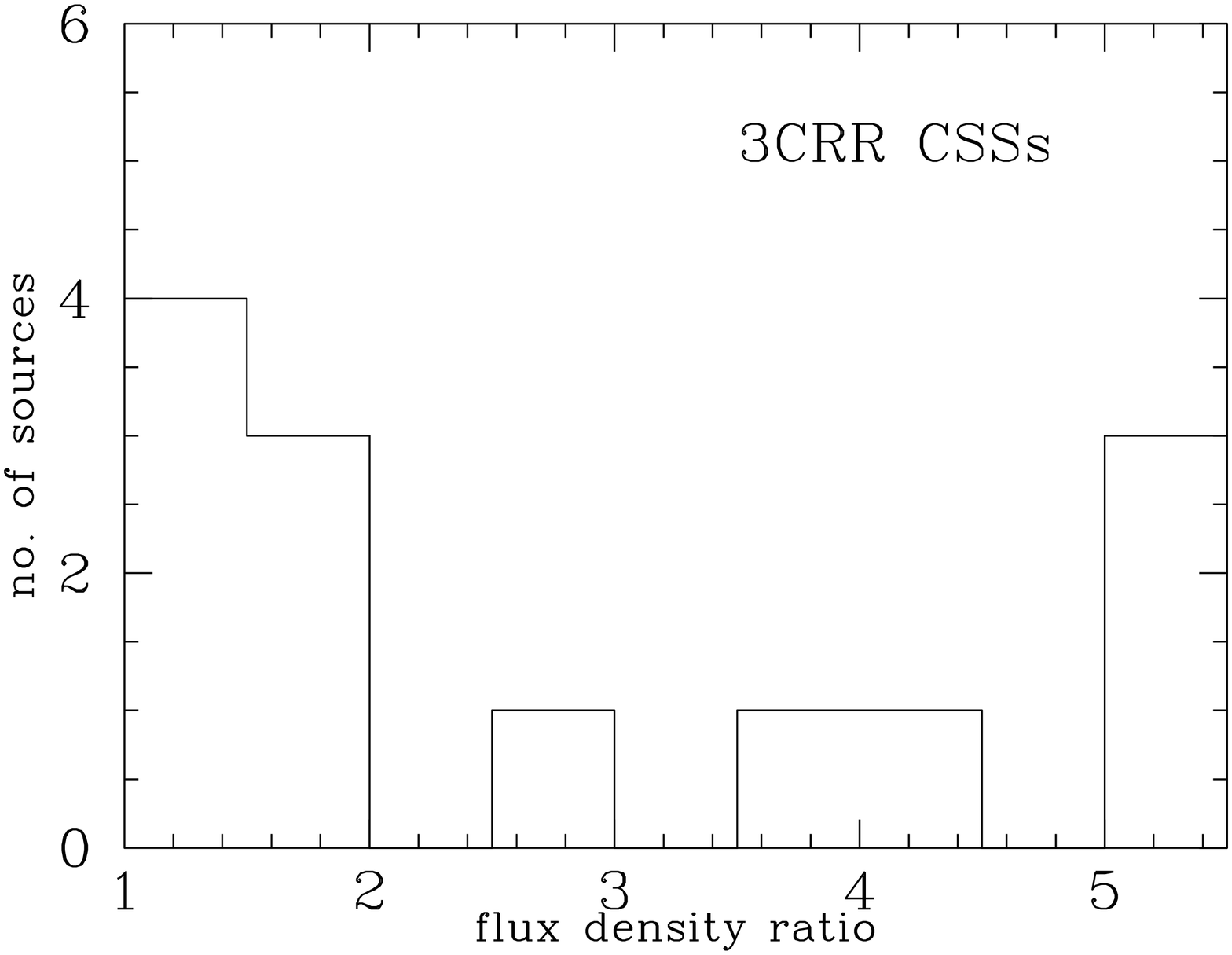,width=1.75in,angle=-90}
   }
}
\caption[]
{The distributions of the flux density
ratio, r$_{\rm L}$ for the B2 and 3CRR sources discussed in the
text. All the sources with r$_{\rm L}$ $>$5 have been
placed in the last bin.
}
\end{figure}

\vspace{0.1cm} \noindent {\bf The 3CRR sample:} It is of interest to compare this with earlier studies of the flux
density ratio for sources in the well-known 3CRR sample. For the 3CRR sources studied by Saikia et al. (2001),
which were confined to the high-luminosity FR\,II sources, the median values of r$_{\rm L}$ for the CSS sources
and the large sources are $\sim$2 and 1.8 respectively (Figure 4, right). This is similar to the values for the B2
sources. The distribution for the CSS sources again appears to have a number of very asymmetric sources with three
of the 13 sources ($\sim$ 23\%) having a value of r$_{\rm L}$$>$5, whereas only five of the 75 ($\sim$7\%) non-CSS
3CRR sources exhibit such a high degree of asymmetry.

\noindent {\bf The B3 CSS sources:} The above result has also been compared with that derived from a sample of CSS
sources selected from the B3--VLA sample with S$_{408}$ $\geq$ 0.8\,Jy (Fanti et al. 2001). Considering all the
doubles, plus the collinear triple, 0744+464, which has a weak central component, the median value of r$_{\rm L}$
for the B3--VLA CSS sample is $\sim$2.4, with eight of the 39 objects ($\sim$21\%) having a value of r$_{\rm L}$
$>$ 5 (Figure 5, left). This is consistent with the results for the B2 and 3CRR samples.

Considering the B2, 3CRR, and B3--VLA samples together, a Kolmogorov-Smirnov test shows the distributions for the
CSS and larger sources to be different at a confidence level of greater than 99\%. However, if the most asymmetric
objects, i.e., those with r$_{\rm L}$$>$5, are excluded, the distributions for the rest of the sources are not
significantly different. The fraction of such asymmetric sources for the combined sample is 25\% for the CSS
sources and 6\% for the larger sources. The increased fraction for the CSS sources is possibly due to interaction
of one of the jets with a dense cloud on one side of the nucleus. These density asymmetries might be intimately
related to the infall of gas which fuels the radio source. Assuming the typical distances of these clouds of gas
from the nucleus of the galaxy to be about 5 to 10\,kpc, the covering factor estimated from the fraction of very
asymmetric sources suggests that the sizes of the `clouds' are similar to those of dwarf galaxies (e.g. Swaters
1999). There could, of course, be a larger number of smaller clouds. However, if the clouds are too small or are
of relatively small mass, they are unlikely to significantly affect the propagation of the jet (e.g. Wang, Wiita,
\& Hooda 2000).

\vspace{0.1cm} \noindent {\bf The flux density ratio--redshift diagram:} It is also of interest to enquire whether
the brightness asymmetries of CSS sources depend on cosmic epoch because of a larger incidence of interactions and
mergers in the past (Ellis et al. 2000).  The r$_{\rm L}$--redshift diagram for the 43 CSS sources with measured
or estimated redshifts from the combined sample (Figure 5, right) shows no significant dependence on redshift.
This suggests that although interactions and  mergers may increase with redshift globally, the flux density
asymmetries in the CSS objects, which are possibly young and still being fuelled by the infall of gas, are similar
at different redshifts.

\begin{figure}
\hspace{1.6cm}
\hbox{
  \psfig{file=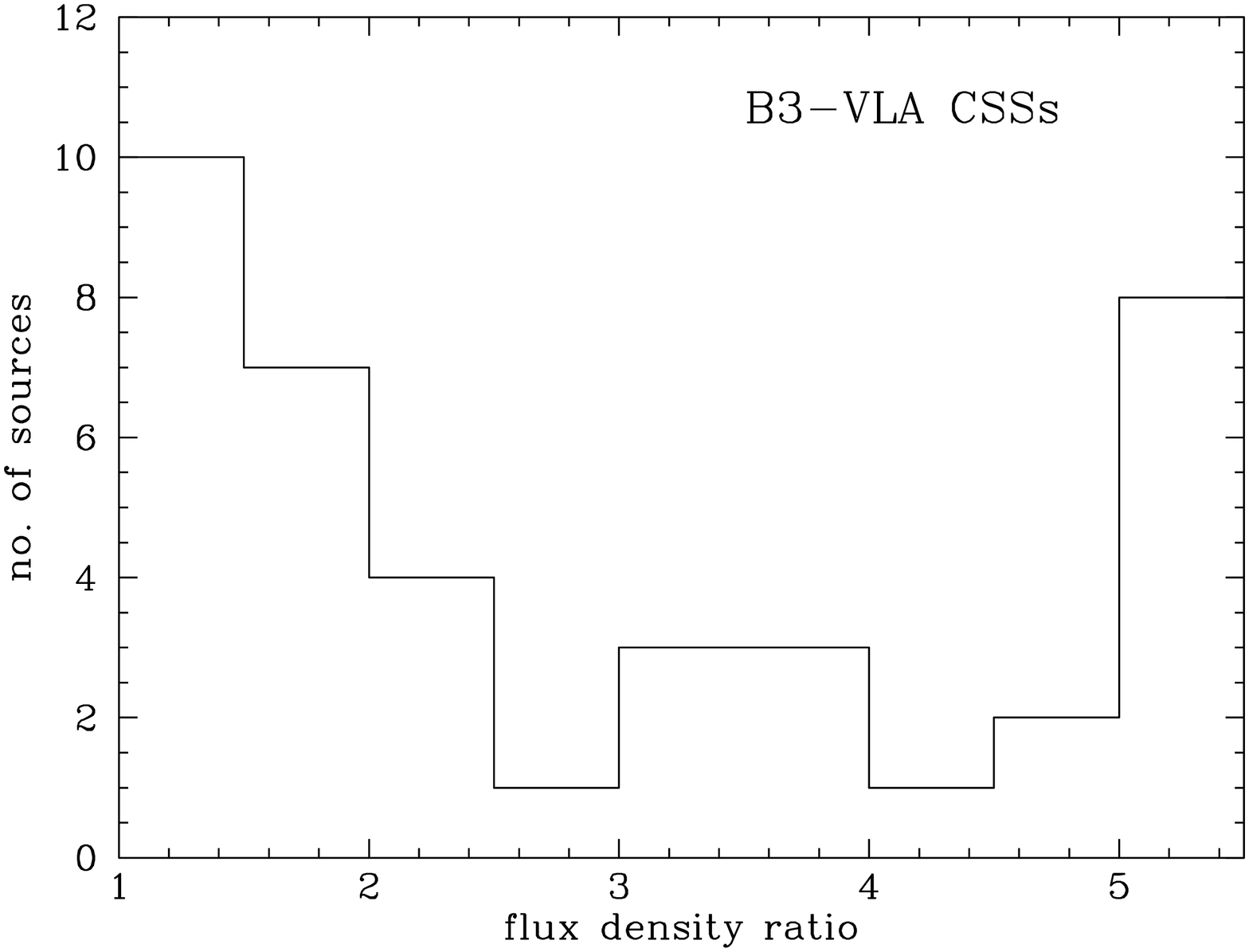,width=2.2in,angle=-90}
  \psfig{file=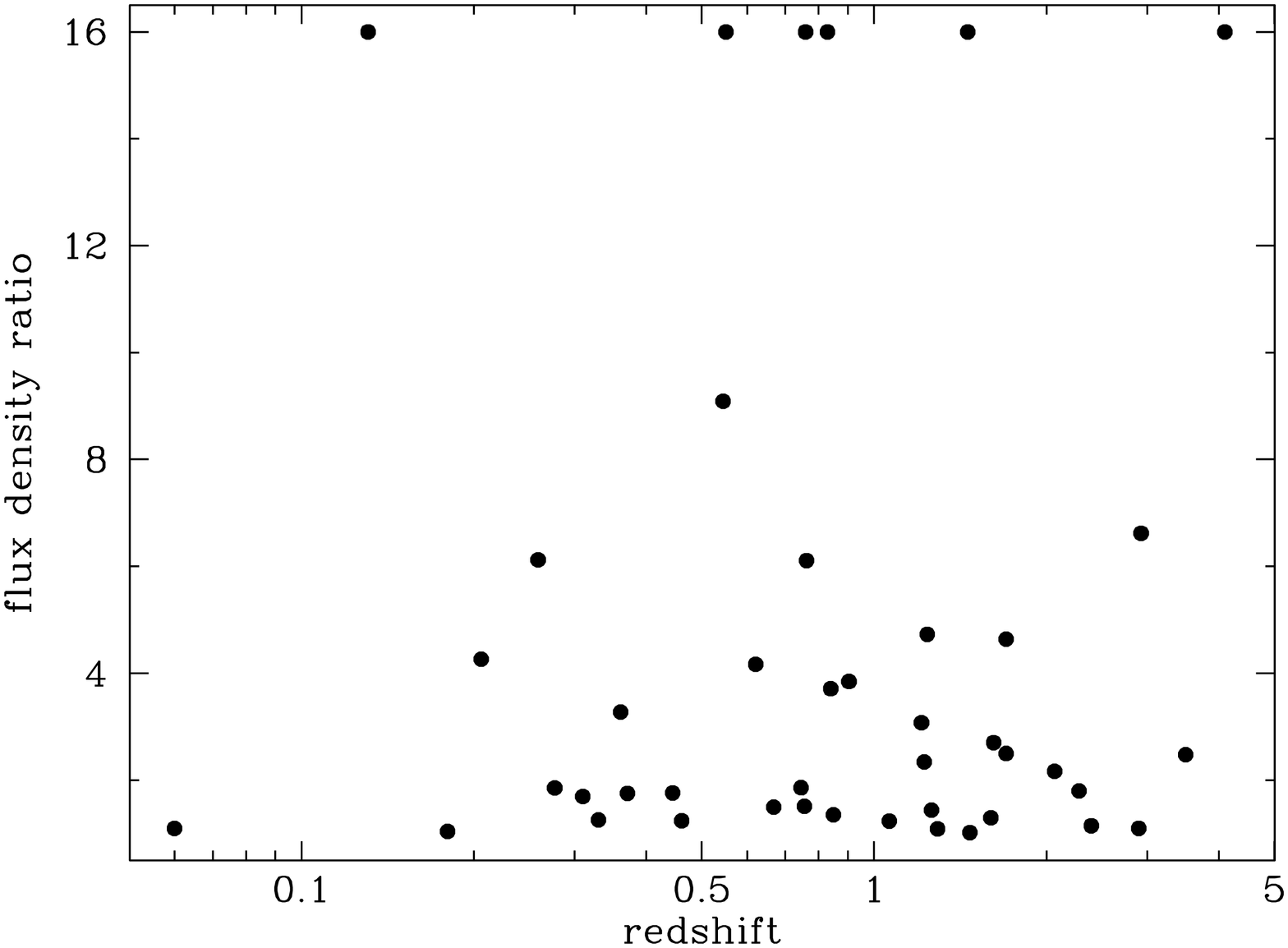,width=2.2in,angle=-90}
   }
\caption[]{Left: The distribution of r$_{\rm L}$ for the B3--VLA CSS sources. All the sources with r$_{\rm L}$
$>$5 have been placed in the last bin. Right: The r$_{\rm L}$--redshift diagram for the B2, 3CRR, and B3 CSS
sources discussed in the text.}
\end{figure}

\begin{figure}
\begin{center}
\hbox{
\hspace{1.0in}
  \psfig{file=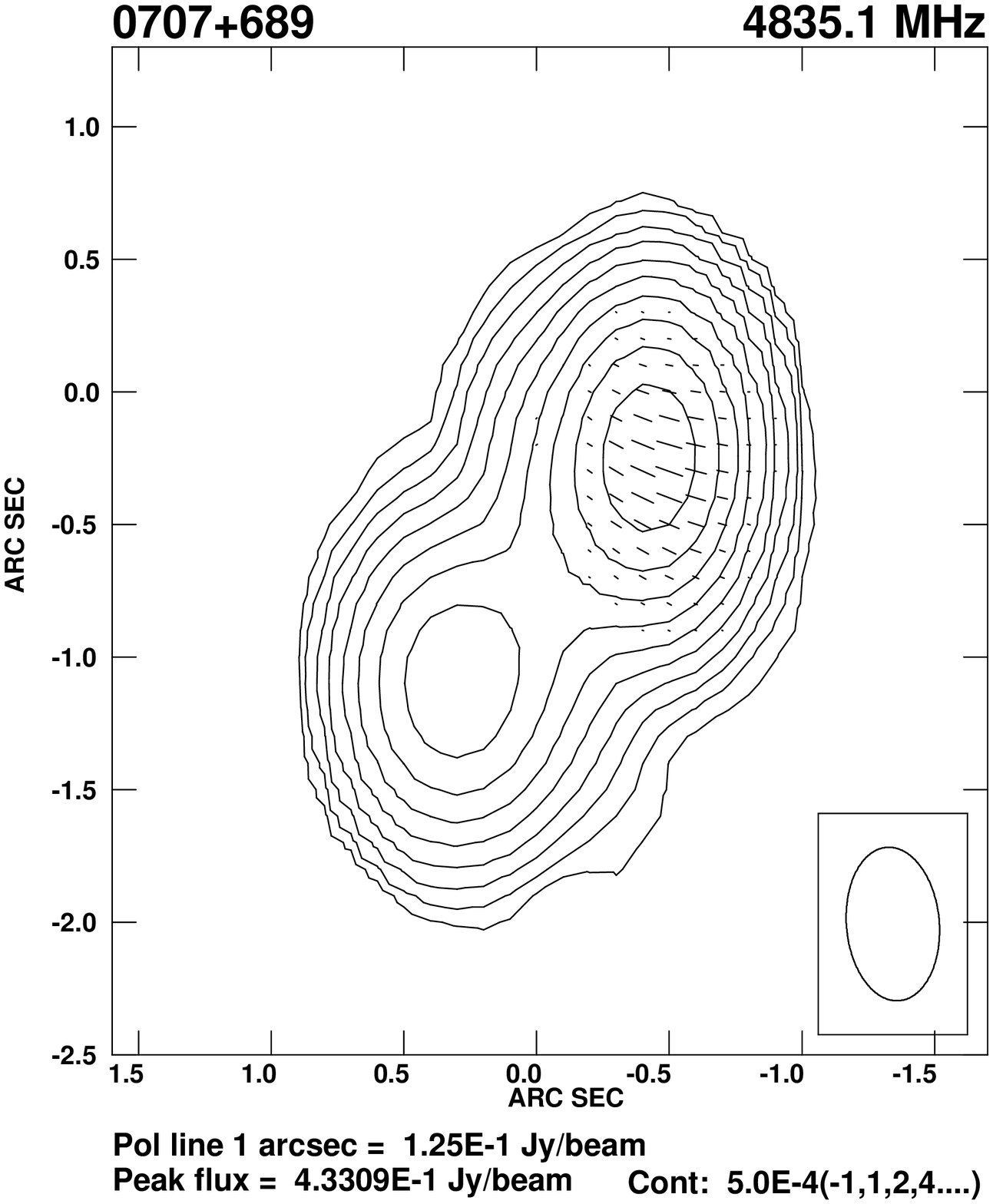,width=2.0in}
\hspace{0.3in}
  \psfig{file=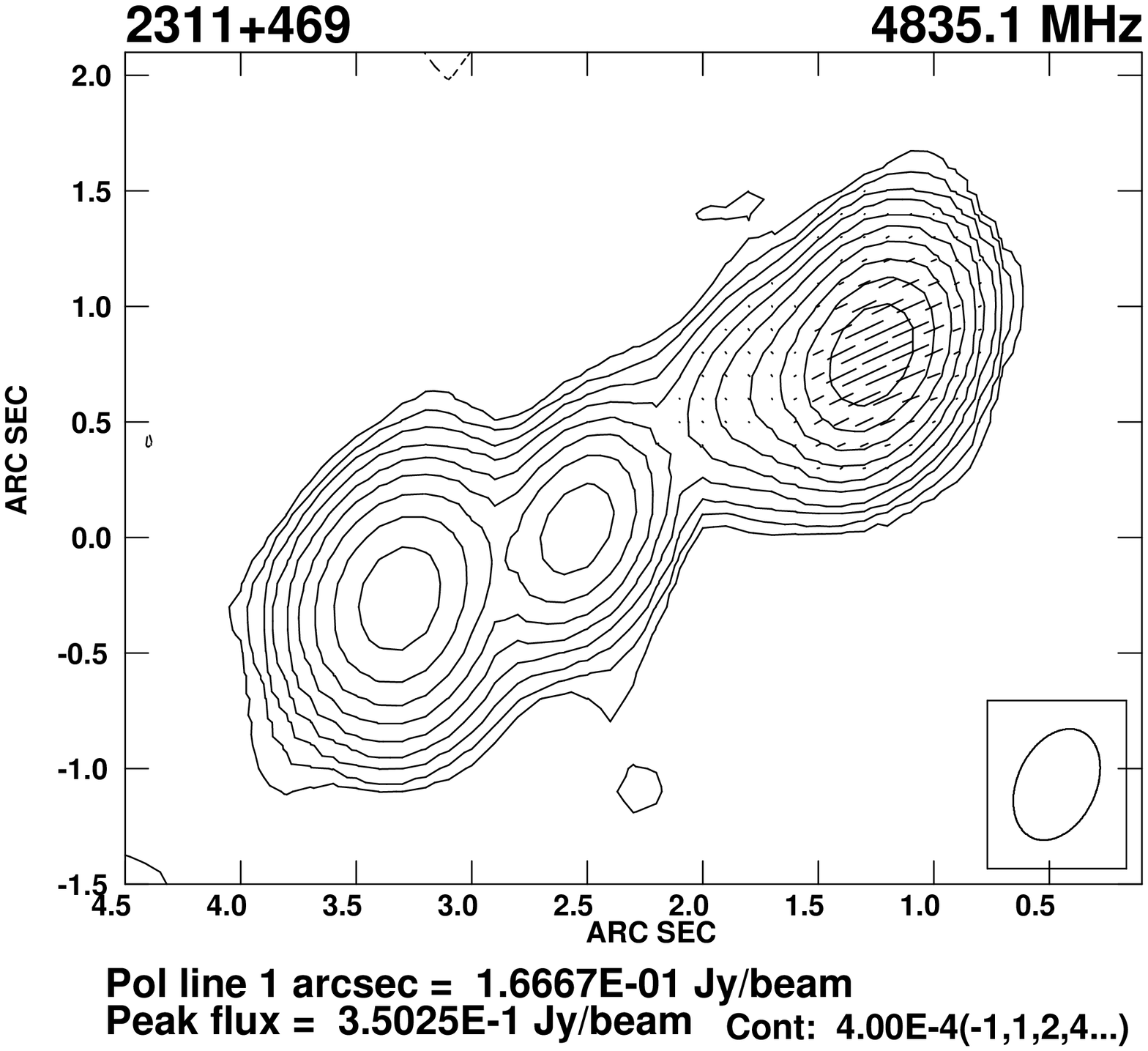,width=2.0in,angle=-90}
} \vspace{-.5cm} \caption[]{VLA A--array images of two CSS objects from the S4 sample, 0707$+$689 and 2311$+$469,
illustrating the high degree of polarisation asymmetry.
}
\end{center}
\vspace{-1.1cm}
\end{figure}

\subsection{Polarisation Measurements}
Evidence of the gas which fuels the radio source may also be probed via polarisation measurements (cf. Saikia \&
Salter 1988; van Breugel et al. 1992; Akujor \& Garrington 1995). In a double-lobed radio source,  the component
interacting with the dense gas should normally both exhibit a higher rotation measure (RM) and also depolarise
more rapidly than the component advancing through the more tenuous medium. Although there have been very few
studies of RM determinations in CSS and GPS objects, partly due to their low levels of polarisation, such studies
have sometimes revealed striking differences in the RM on opposite sides of the nucleus. Detailed polarisation
observations of 3C\,147  show huge differential RMs between the two oppositely directed lobes, suggesting their
evolution in an asymmetric environment (Junor et al.\ 1999). A similar situation is also seen in the quasar
0725$+$147 (Mantovani et al.\ 1994). These asymmetries in the central regions of active galaxies must be
intimately related to the supply of fuel to the central engine in these objects, possibly due to interactions with
companion galaxies.

\begin{table}
\caption{The polarisation parameters} \vspace*{1ex}
\begin{tabular}{c c c c c c c}
Source & Opt. Id. & z &   Resolution & m$_{west}$ & m$_{east}$ &  m$_{west}$/m$_{east}$ \\

B0707+689  &  Q  & 1.141 & 0.$^{\prime\prime}$58$\times$0.$^{\prime\prime}$35 along 5$^{\circ}$ &
             2.61  & $<$0.23   &  $>$11.3   \\
B2311+469  &  Q  & 0.742 & 0.$^{\prime\prime}$51$\times$0.$^{\prime\prime}$34 along 155$^{\circ}$
           & 9.72  & $<$0.20  & $>$48.6 \\
\end{tabular}
\end{table}

In the absence of RM determinations, one can also examine the asymmetry in the polarisation of the two oppositely
directed lobes of radio emission. A sample of CSS sources has been observed with the VLA to determine the
asymmetry in the polarisation of the oppositely-directed lobes (D.J.\ Saikia et al., in preparation). Two of these
sources are presented in Figure 6, and their polarisation parameters are listed in Table 1. The median value of
the ratio of the degree of polarisation of the oppositely-directed lobes at $\lambda$\,6 or 3.6\,cm is $\sim$6 for
the CSS objects compared with $\sim$1.5 for the larger sources. A Kolmogorov-Smirnov test shows the two
populations to be different at a significance level of $>$99.9\%. This high degree of polarisation asymmetry is
also likely to be due to interactions with dense clouds of gas which could either compress magnetic fields,
increasing the degree of polarisation, or else depolarise the emission. The degree of polarisation asymmetry is
much higher than due to the Laing--Garrington effect (Laing 1988; Garrington et al. 1988). In a number of cases,
it is the jet side which is more strongly depolarised (e.g. 3C\,459) or has a much higher RM (e.g. 3C\,147). Also,
the typical core radii of the halos which cause the Laing-Garrington effect is $\sim$100\,kpc, which will cause
only marginal polarisation asymmetry on the scale of the CSS objects (cf. Garrington \& Conway 1991).

\section{Numerical Simulation}
We have attempted to understand the evolution of CSS sources through an asymmetric environment in the nuclear
region by calculating analytical models (Saikia et al.\ 1996) as well as performing two- and three-dimensional
simulations (Figure 7) of the propagation of jets through asymmetric environments (Jeyakumar et al. in
preparation). In this study, intrinsically similar jets with external  Mach numbers of 26.0 propagate through
ambient media with different gas distributions in the inner 5\,kpc; at first the bow shocks run substantially
ahead of the jet working surfaces. The jet propagating through the less dense material naturally emerges faster,
and the value of r$_{\rm D}$ rises quickly; it later declines as the jets traverse identical media further out
(Figure 8).  The numerical simulations qualitatively agree with the analytical models  and support the hypothesis
that CSS sources show greater asymmetries than do FR\,IIs predominantly because ambient gas asymmetries (the AGN
fuel?) are concentrated in the inner portions of the galaxies.

\begin{figure}
\begin{center}
\hspace{.4cm}
\hbox{
\psfig{file=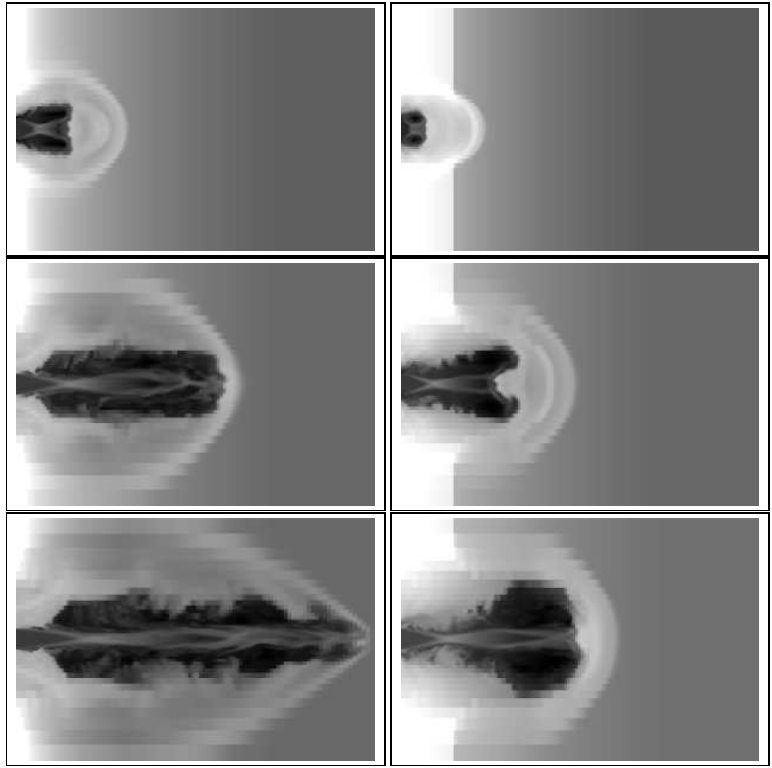,width=4.0in}
}
\caption[]
{The logarithm of the density (with white highest and black lowest)
is shown for  3-D simulations with identical
jet parameters passing through
a smooth decline in external gas density (left) and a more gradual decline, followed by
an abrupt drop to an identical gas density after 5 kpc (right).  The times
correspond to $1.3 \times 10^5$yr (top),
$4.8  \times 10^5$yr (middle), and $8.6 \times 10^5$yr (bottom).
}
\end{center}
\end{figure}

\begin{figure*}
\vspace{-2.7in}
\begin{center}
\hbox{
\psfig{file=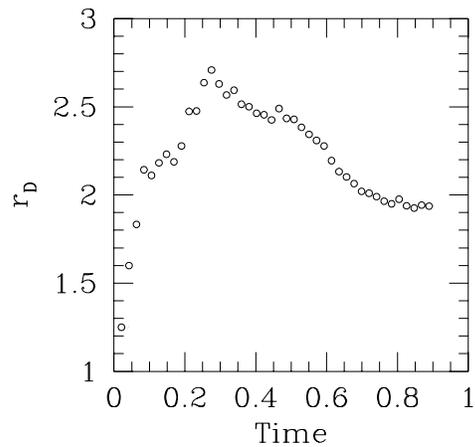,width=\textwidth,angle=-90}
}
\end{center}
\vspace{-1in}
\caption[]{The variation of $r_{\rm D}$ estimated using the Mach disk as a function of
time scaled to units of Myr.}
\end{figure*}



%
%







\section*{References}






\reference Akujor, C.E., \& Garrington, S.T.\ 1995, A\&AS, 112, 235

\reference Arshakian, T.G., \&  Longair, M.S.\ 2000, MNRAS, 311, 846

\reference Bremer, M.N.\ 1997, MNRAS, 284, 126

\reference Carvalho, J.C.\ 1985, MNRAS, 215, 463

\reference Colla, G., et al.\ 1970, A\&AS, 1, 281

\reference Crawford, C.S., \& Vanderriest, C.\ 1997, MNRAS, 285, 580

\reference Dallacasa, D., Fanti, C., Fanti, R., Schilizzi, R.T., \& Spencer, R.E.\
           1995, A\&A, 295, 27

\reference De Young, D.S.\ 1997, ApJ, 490, 55L

\reference Ellis, R.S., Abraham, R.G., Brinchmann, J., \& Menanteau, F.\ 2000, A\&G, No. 2, 10

\reference Fanti, C., Fanti, R., Dallacasa, D., Schilizzi, R.T., Spencer, R.E., \&
           Stanghellini, C.\ 1995, A\&A, 302, 317

\reference Fanti, C., Pozzi, F., Dallacasa, D., Fanti, R., Gregorini, L., Stanghellini, C., \& Vigotti, M.\ 2001,
A\&A, 369, 380

\reference Garrington, S.T., \& Conway, R.G.\ 1991, MNRAS, 250, 198

\reference Garrington, S.T., Leahy, J.P., Conway, R.G., \& Laing, R.A.\
           1988, Nature, 331, 147

\reference Hernquist, L., \& Mihos, J.C.\ 1995, ApJ, 448, 41

\reference Jeyakumar, S., \& Saikia, D.J.\ 2000, MNRAS, 311, 397

\reference Junor, W., Salter, C., Saikia, D., Mantovani, F., \& Peck, A.\
           1999, MNRAS, 308, 955

\reference Laing, R.A.\ 1988, Nature, 331, 149

\reference Mantovani, F., Junor, W., Fanti, R., Padrielli, L., \& Saikia, D.J.\
           1994, A\&A, 292, 59

\reference Mantovani, F., Saikia, D.J., Browne, I.W.A., Fanti, R., Muxlow, T.W.B., \&
           Padrielli, L.\ 1990, MNRAS, 245, 427

\reference McCarthy, P.J., van Breugel, W., \& Kapahi, V.K.\ 1991, ApJ, 371, 478

\reference Murgia, M., Fanti, C., Fanti, R., Gregorini, L., Klein, U., Mack, K.-H., \& Vigotti, M.\ 1999, A\&A,
345, 769

\reference O'Dea, C.P.\ 1998, PASP, 110, 493

\reference Owen, F.N., \& Puschell, J.J.\ 1984, AJ, 89, 932

\reference Owsianik, I., \& Conway, J.E.\ 1998, A\&A, 337, 69

\reference Perucho, M., \& Mart\'{i}, J.M.\ 2002, ApJ, 568, 639

\reference Readhead, A., Taylor, G., Pearson, T., \& Wilkinson, P.\
           1996, ApJ, 460, 634

\reference Saikia, D.J., \& Salter, C.J.,\ 1988, ARA\&A, 26, 93

\reference Saikia, D.J., Salter, C.J., \& Muxlow, T.W.B.\ 1987, MNRAS, 224, 911

\reference Saikia, D.J., Jeyakumar, S., Wiita, P.J.,  \& Hooda, J.S.\ 1996 in
           The Second Workshop on GPS and CSS Sources,
           ed.\ I.A.G. Snellen et al. (Leiden: Leiden Observatory), 252

\reference Saikia, D.J., Jeyakumar, S., Salter, C.J., Thomasson, P., Spencer, R.E., \& Mantovani, F.\ 2001, MNRAS,
321, 37

\reference Saikia, D.J., Thomasson, P., Spencer, R.E., Mantovani, F., Salter, C.J., \& Jeyakumar, S.\ 2002, A\&A,
391, 149

\reference Snellen, I.A.G., Schilizzi, R.T., Miley, G.K., de Bruyn, A.G.,
           Bremer, M.N., \& R\"{o}ttgering, H.J.A.\ 2000, MNRAS, 319, 445

\reference Swaters, R.A.\ 1999, PhD thesis, Rijksuniversiteit Groningen

\reference Tadhunter C., Dickson R., Morganti R., Robinson, T.G., Wills, K., Villar-Martin, M., \& Hughes, M.\
2002, MNRAS, 330, 997

\reference Taylor, G.B., Marr, J.M., Pearson, T.J., \& Readhead, A.C.S.\
           2000, ApJ, 541, 112

\reference Thomasson, P., Saikia, D.J., \& Muxlow, T.W.B.\ 2002, MNRAS, 341, 91

\reference van Breugel, W.J.M., Fanti, C., Fanti, R., Stanghellini, C., Schilizzi, R.T., \&
           Spencer, R.E.\ 1992, A\&A, 256, 56

\reference Wang, Z., Wiita, P.J., \& Hooda, J.S.\ 2000, ApJ,  534, 201


\end{document}